\documentclass[conference,a4paper]{IEEEtran}
\IEEEoverridecommandlockouts
\usepackage[acronyms,nonumberlist,nopostdot,nomain,nogroupskip,acronymlists={hidden}]{glossaries}
\newglossary[algh]{hidden}{acrh}{acnh}{Hidden Acronyms}

\usepackage{cite}
\usepackage{amsmath,amssymb,amsfonts}
\usepackage{algorithmic}
\usepackage{textcomp}
\usepackage{xcolor}
\usepackage{booktabs}
\usepackage{siunitx}

\usepackage{booktabs}
\usepackage{makecell}

\usepackage{url}
\usepackage{pgfplots}
\pgfplotsset{compat=newest}
\pgfplotsset{plot coordinates/math parser=false}
\newlength\fheight
\newlength\fwidth
\usepgfplotslibrary{patchplots,groupplots}
\pgfplotsset{compat=1.18}

\let\svthefootnote\thefootnote
\newcommand{\unnumberedfootnote}[1]{%
    \let\thefootnote\relax\footnotetext{#1}%
    \let\thefootnote\svthefootnote%
}

\def\BibTeX{{\rm B\kern-.05em{\sc i\kern-.025em b}\kern-.08em
    T\kern-.1667em\lower.7ex\hbox{E}\kern-.125emX}}

\expandafter\def\csname ver@fixltx2e.sty\endcsname{}

\usepackage[font=footnotesize]{caption}
\usepackage{subcaption}
\usepackage{subcaption} 
\usepackage{soul}

\usepackage{xspace}
\newcommand{\project}{SioLENA\xspace}

\newacronym{dt}{DT}{Digital Twin}
\newacronym{3gpp}{3GPP}{3rd Generation Partnership Project}
\newacronym{3d}{3D}{three dimensional}
\newacronym{4g}{4G}{4th generation}
\newacronym{5g}{5G}{5th generation}
\newacronym{6g}{6G}{6th generation}
\newacronym{5gc}{5GC}{5G Core}
\newacronym{adc}{ADC}{Analog to Digital Converter}
\newacronym{aerpaw}{AERPAW}{Aerial Experimentation and Research Platform for Advanced Wireless}
\newacronym{ai}{AI}{Artificial Intelligence}
\newacronym{aimd}{AIMD}{Additive Increase Multiplicative Decrease}
\newacronym{am}{AM}{Acknowledged Mode}
\newacronym{amc}{AMC}{Adaptive Modulation and Coding}
\newacronym{amf}{AMF}{Access and Mobility Management Function}
\newacronym{aops}{AOPS}{Adaptive Order Prediction Scheduling}
\newacronym{api}{API}{Application Programming Interface}
\newacronym{apn}{APN}{Access Point Name}
\newacronym{ap}{AP}{Application Protocol}
\newacronym{aqm}{AQM}{Active Queue Management}
\newacronym{ausf}{AUSF}{Authentication Server Function}
\newacronym{avc}{AVC}{Advanced Video Coding}
\newacronym{awgn}{AGWN}{Additive White Gaussian Noise}
\newacronym{balia}{BALIA}{Balanced Link Adaptation Algorithm}
\newacronym{bbu}{BBU}{Base Band Unit}
\newacronym{bdp}{BDP}{Bandwidth-Delay Product}
\newacronym{ber}{BER}{Bit Error Rate}
\newacronym{bf}{BF}{Beamforming}
\newacronym{bler}{BLER}{Block Error Rate}
\newacronym{brr}{BRR}{Bayesian Ridge Regressor}
\newacronym{bs}{BS}{Base Station}
\newacronym{bsr}{BSR}{Buffer Status Report}
\newacronym{bss}{BSS}{Business Support System}
\newacronym{ca}{CA}{Carrier Aggregation}
\newacronym{caas}{CaaS}{Connectivity-as-a-Service}
\newacronym{cb}{CB}{Code Block}
\newacronym{cc}{CC}{Congestion Control}
\newacronym{ccid}{CCID}{Congestion Control ID}
\newacronym{cco}{CC}{Carrier Component}
\newacronym{cdd}{CDD}{Cyclic Delay Diversity}
\newacronym{cdf}{CDF}{Cumulative Distribution Function}
\newacronym{cdn}{CDN}{Content Distribution Network}
\newacronym{cir}{CIR}{Channel Impulse Response}
\newacronym{cli}{CLI}{Command-line Interface}
\newacronym{cn}{CN}{Core Network}
\newacronym{cnn}{CNN}{Convolutional Neural Network}
\newacronym{codel}{CoDel}{Controlled Delay Management}
\newacronym{comac}{COMAC}{Converged Multi-Access and Core}
\newacronym{cord}{CORD}{Central Office Re-architected as a Datacenter}
\newacronym{cornet}{CORNET}{COgnitive Radio NETwork}
\newacronym{cosmos}{COSMOS}{Cloud Enhanced Open Software Defined Mobile Wireless Testbed for City-Scale Deployment}
\newacronym{cots}{COTS}{Commercial Off-the-Shelf}
\newacronym{cp}{CP}{Control Plane}
\newacronym{cyp}{CP}{Cyclic Prefix}
\newacronym{up}{UP}{User Plane}
\newacronym{cpu}{CPU}{Central Processing Unit}
\newacronym{cqi}{CQI}{Channel Quality Information}
\newacronym{cql}{CQL}{Conservative Q-Learning}
\newacronym{cr}{CR}{Cognitive Radio}
\newacronym{cran}{CRAN}{Cloud \gls{ran}}
\newacronym{crs}{CRS}{Cell Reference Signal}
\newacronym{csi}{CSI}{Channel State Information}
\newacronym{csirs}{CSI-RS}{Channel State Information - Reference Signal}
\newacronym{cu}{CU}{Central Unit}
\newacronym{cucp}{CU-CP}{Central Unit Control Plane}
\newacronym{cuup}{CU-UP}{Central Unit User Plane}
\newacronym{d2tcp}{D$^2$TCP}{Deadline-aware Data center TCP}
\newacronym{d3}{D$^3$}{Deadline-Driven Delivery}
\newacronym{dac}{DAC}{Digital to Analog Converter}
\newacronym{dag}{DAG}{Directed Acyclic Graph}
\newacronym{das}{DAS}{Distributed Antenna System}
\newacronym{dash}{DASH}{Dynamic Adaptive Streaming over HTTP}
\newacronym{dc}{DC}{Dual Connectivity}
\newacronym{dccp}{DCCP}{Datagram Congestion Control Protocol}
\newacronym{dce}{DCE}{Direct Code Execution}
\newacronym{dci}{DCI}{Downlink Control Information}
\newacronym{dctcp}{DCTCP}{Data Center TCP}
\newacronym{dl}{DL}{Downlink}
\newacronym{dmr}{DMR}{Deadline Miss Ratio}
\newacronym{dmrs}{DMRS}{DeModulation Reference Signal}
\newacronym{dqn}{DQN}{Deep Q-Network}
\newacronym{drlcc}{DRL-CC}{Deep Reinforcement Learning Congestion Control}
\newacronym{drs}{DRS}{Discovery Reference Signal}
\newacronym{du}{DU}{Distributed Unit}
\newacronym{ee}{EE}{Energy Efficiency}
\newacronym{e2e}{E2E}{end-to-end}
\newacronym{earfcn}{EARFCN}{E-UTRA Absolute Radio Frequency Channel Number}
\newacronym{ecaas}{ECaaS}{Edge-Cloud-as-a-Service}
\newacronym{ecn}{ECN}{Explicit Congestion Notification}
\newacronym{edf}{EDF}{Earliest Deadline First}
\newacronym{embb}{eMBB}{Enhanced Mobile Broadband}
\newacronym{empower}{EMPOWER}{EMpowering transatlantic PlatfOrms for advanced WirEless Research}
\newacronym{enb}{eNB}{evolved Node Base}
\newacronym{epc}{EPC}{Evolved Packet Core}
\newacronym{eps}{EPS}{Evolved Packet System}
\newacronym{es}{ES}{Edge Server}
\newacronym{etsi}{ETSI}{European Telecommunications Standards Institute}
\newacronym[firstplural=Estimated Times of Arrival (ETAs)]{eta}{ETA}{Estimated Time of Arrival}
\newacronym{eutran}{E-UTRAN}{Evolved Universal Terrestrial Access Network}
\newacronym{faas}{FaaS}{Function-as-a-Service}
\newacronym{fapi}{FAPI}{Functional Application Platform Interface}
\newacronym{fdd}{FDD}{Frequency Division Duplexing}
\newacronym{fdm}{FDM}{Frequency Division Multiplexing}
\newacronym{fdma}{FDMA}{Frequency Division Multiple Access}
\newacronym{fed4fire}{FED4FIRE+}{Federation 4 Future Internet Research and Experimentation Plus}
\newacronym{fir}{FIR}{Finite Impulse Response}
\newacronym{fit}{FIT}{Future \acrlong{iot}}
\newacronym{fpga}{FPGA}{Field Programmable Gate Array}
\newacronym{fr1}{FR1}{Frequency Range 1}
\newacronym{fr2}{FR2}{Frequency Range 2}
\newacronym{fr3}{FR3}{Frequency Range 3}
\newacronym{fs}{FS}{Fast Switching}
\newacronym{fscc}{FSCC}{Flow Sharing Congestion Control}
\newacronym{ftp}{FTP}{File Transfer Protocol}
\newacronym{fw}{FW}{Flow Window}
\newacronym{ge}{GE}{Gaussian Elimination}
\newacronym{gnb}{gNB}{Next Generation Node Base}
\newacronym{gop}{GOP}{Group of Pictures}
\newacronym{gpr}{GPR}{Gaussian Process Regressor}
\newacronym{gpu}{GPU}{Graphics Processing Unit}
\newacronym{gtp}{GTP}{GPRS Tunneling Protocol}
\newacronym{gtpc}{GTP-C}{GPRS Tunnelling Protocol Control Plane}
\newacronym{gtpu}{GTP-U}{GPRS Tunnelling Protocol User Plane}
\newacronym{gtpv2c}{GTPv2-C}{\gls{gtp} v2 - Control}
\newacronym{gw}{GW}{Gateway}
\newacronym{harq}{HARQ}{Hybrid Automatic Repeat reQuest}
\newacronym{hetnet}{HetNet}{Heterogeneous Network}
\newacronym{hh}{HH}{Hard Handover}
\newacronym{hol}{HOL}{Head-of-Line}
\newacronym{hqf}{HQF}{Highest-quality-first}
\newacronym{hss}{HSS}{Home Subscription Server}
\newacronym{http}{HTTP}{HyperText Transfer Protocol}
\newacronym{ia}{IA}{Initial Access}
\newacronym{iab}{IAB}{Integrated Access and Backhaul}
\newacronym{ic}{IC}{Incident Command}
\newacronym{ietf}{IETF}{Internet Engineering Task Force}
\newacronym{ieee}{IEEE}{Institute of Electrical and Electronics Engineers
}
\newacronym{imsi}{IMSI}{International Mobile Subscriber Identity}
\newacronym{imt}{IMT}{International Mobile Telecommunication}
\newacronym{iot}{IoT}{Internet of Things}
\newacronym{ip}{IP}{Internet Protocol}
\newacronym{itu}{ITU}{International Telecommunication Union}
\newacronym{kpi}{KPI}{Key Performance Indicator}
\newacronym{kpm}{KPM}{Key Performance Measurement}
\newacronym{kvm}{KVM}{Kernel-based Virtual Machine}
\newacronym{los}{LoS}{Line-of-Sight}
\newacronym{lsm}{LSM}{Link-to-System Mapping}
\newacronym{lstm}{LSTM}{Long Short Term Memory}
\newacronym{lte}{LTE}{Long Term Evolution}
\newacronym{lxc}{LXC}{Linux Container}
\newacronym{m2m}{M2M}{Machine to Machine}
\newacronym{mac}{MAC}{Medium Access Control}
\newacronym{manet}{MANET}{Mobile Ad Hoc Network}
\newacronym{mano}{MANO}{Management and Orchestration}
\newacronym{mc}{MC}{Multi-Connectivity}
\newacronym{mcc}{MCC}{Mobile Cloud Computing}
\newacronym{mchem}{MCHEM}{Massive Channel Emulator}
\newacronym{mcs}{MCS}{Modulation and Coding Scheme}
\newacronym{mec}{MEC}{Multi-access Edge Computing}
\newacronym{mec2}{MEC}{Mobile Edge Cloud}
\newacronym{mfc}{MFC}{Mobile Fog Computing}
\newacronym{mgen}{MGEN}{Multi-Generator}
\newacronym{mi}{MI}{Mutual Information}
\newacronym{mib}{MIB}{Master Information Block}
\newacronym{miesm}{MIESM}{Mutual Information Based Effective SINR}
\newacronym{mimo}{MIMO}{Multiple Input, Multiple Output}
\newacronym{ml}{ML}{Machine Learning}
\newacronym{mlr}{MLR}{Maximum-local-rate}
\newacronym[plural=\gls{mme}s,firstplural=Mobility Management Entities (MMEs)]{mme}{MME}{Mobility Management Entity}
\newacronym{mmtc}{mMTC}{Massive Machine-Type Communications}
\newacronym{mmwave}{mmWave}{millimeter wave}
\newacronym{mpdccp}{MP-DCCP}{Multipath Datagram Congestion Control Protocol}
\newacronym{mptcp}{MPTCP}{Multipath TCP}
\newacronym{mr}{MR}{Maximum Rate}
\newacronym{mrdc}{MR-DC}{Multi \gls{rat} \gls{dc}}
\newacronym{mse}{MSE}{Mean Square Error}
\newacronym{mss}{MSS}{Maximum Segment Size}
\newacronym{mt}{MT}{Mobile Termination}
\newacronym{mtd}{MTD}{Machine-Type Device}
\newacronym{mtu}{MTU}{Maximum Transmission Unit}
\newacronym{mumimo}{MU-MIMO}{Multi-user \gls{mimo}}
\newacronym{mvno}{MVNO}{Mobile Virtual Network Operator}
\newacronym{nalu}{NALU}{Network Abstraction Layer Unit}
\newacronym{nas}{NAS}{Network Attached Storage}
\newacronym{nat}{NAT}{Network Address Translation}
\newacronym{nbiot}{NB-IoT}{Narrow Band IoT}
\newacronym{nfv}{NFV}{Network Function Virtualization}
\newacronym{nfvi}{NFVI}{Network Function Virtualization Infrastructure}
\newacronym{ni}{NI}{Network Interfaces}
\newacronym{nic}{NIC}{Network Interface Card}
\newacronym{nist}{NIST}{National Institute of Standards and Technolog}
\newacronym{nlos}{NLoS}{Non-Line-of-Sight}
\newacronym{now}{NOW}{Non Overlapping Window}
\newacronym{nsm}{NSM}{Network Service Mesh}
\newacronym{nr}{NR}{New Radio}
\newacronym{nextg}{NextG}{Next Generation}
\newacronym{nrf}{NRF}{Network Repository Function}
\newacronym{nsa}{NSA}{Non Stand Alone}
\newacronym{nse}{NSE}{Network Slicing Engine}
\newacronym{nssf}{NSSF}{Network Slice Selection Function}
\newacronym{o2i}{O2I}{Outdoor to Indoor}
\newacronym{oai}{OAI}{OpenAirInterface}
\newacronym{oaicn}{OAI-CN}{\gls{oai} \acrlong{cn}}
\newacronym{oairan}{OAI-RAN}{\acrlong{oai} \acrlong{ran}}
\newacronym{oam}{OAM}{Operations, Administration and Maintenance}
\newacronym{ofdm}{OFDM}{Orthogonal Frequency Division Multiplexing}
\newacronym{olia}{OLIA}{Opportunistic Linked Increase Algorithm}
\newacronym{omec}{OMEC}{Open Mobile Evolved Core}
\newacronym{onap}{ONAP}{Open Network Automation Platform}
\newacronym{onf}{ONF}{Open Networking Foundation}
\newacronym{onos}{ONOS}{Open Networking Operating System}
\newacronym{oom}{OOM}{\gls{onap} Operations Manager}
\newacronym{opnfv}{OPNFV}{Open Platform for \gls{nfv}}
\newacronym[type=hidden]{oran}{O-RAN}{Open \gls{ran}}
\newacronym{orbit}{ORBIT}{Open-Access Research Testbed for Next-Generation Wireless Networks}
\newacronym{os}{OS}{Operating System}
\newacronym{osm2}{OSM}{Open Street Map}
\newacronym{oss}{OSS}{Operations Support System}
\newacronym{p2mp}{P2MP}{Point-to-multipoint}
\newacronym{pa}{PA}{Position-aware}
\newacronym{pase}{PASE}{Prioritization, Arbitration, and Self-adjusting Endpoints}
\newacronym{pawr}{PAWR}{Platforms for Advanced Wireless Research}
\newacronym{pbch}{PBCH}{Physical Broadcast Channel}
\newacronym{pcef}{PCEF}{Policy and Charging Enforcement Function}
\newacronym{pcfich}{PCFICH}{Physical Control Format Indicator Channel}
\newacronym{pcrf}{PCRF}{Policy and Charging Rules Function}
\newacronym{pdcch}{PDCCH}{Physical Downlink Control Channel}
\newacronym{pdcp}{PDCP}{Packet Data Convergence Protocol}
\newacronym{pdsch}{PDSCH}{Physical Downlink Shared Channel}
\newacronym{pdu}{PDU}{Packet Data Unit}
\newacronym{pf}{PF}{Proportional Fair}
\newacronym{pgw}{PGW}{Packet Gateway}
\newacronym{phich}{PHICH}{Physical Hybrid ARQ Indicator Channel}
\newacronym{phy}{PHY}{Physical}
\newacronym{pl}{PL}{Path Loss}
\newacronym{pmch}{PMCH}{Physical Multicast Channel}
\newacronym{pmi}{PMI}{Precoding Matrix Indicators}
\newacronym{powder}{POWDER}{Platform for Open Wireless Data-driven Experimental Research}
\newacronym{ppo}{PPO}{Proximal Policy Optimization}
\newacronym{ppp}{PPP}{Poisson Point Process}
\newacronym{prach}{PRACH}{Physical Random Access Channel}
\newacronym{prb}{PRB}{Physical Resource Block}
\newacronym{psnr}{PSNR}{Peak Signal to Noise Ratio}
\newacronym{pss}{PSS}{Primary Synchronization Signal}
\newacronym{pucch}{PUCCH}{Physical Uplink Control Channel}
\newacronym{pusch}{PUSCH}{Physical Uplink Shared Channel}
\newacronym{qam}{QAM}{Quadrature Amplitude Modulation}
\newacronym{qci}{QCI}{\gls{qos} Class Identifier}
\newacronym{qoe}{QoE}{Quality of Experience}
\newacronym{qos}{QoS}{Quality of Service}
\newacronym{quic}{QUIC}{Quick UDP Internet Connections}
\newacronym{rach}{RACH}{Random Access Channel}
\newacronym{ran}{RAN}{Radio Access Network}
\newacronym[firstplural=Radio Access Technologies (RATs)]{rat}{RAT}{Radio Access Technology}
\newacronym{rbg}{RBG}{Resource Block Group}
\newacronym{rcn}{RCN}{Research Coordination Network}
\newacronym{rc}{RC}{RAN Control}
\newacronym{rec}{REC}{Radio Edge Cloud}
\newacronym{red}{RED}{Random Early Detection}
\newacronym{renew}{RENEW}{Reconfigurable Eco-system for Next-generation End-to-end Wireless}
\newacronym{rf}{RF}{Radio Frequency}
\newacronym{rfc}{RFC}{Request for Comments}
\newacronym{rfr}{RFR}{Random Forest Regressor}
\newacronym{ric}{RIC}{RAN Intelligent Controller}
\newacronym{rlc}{RLC}{Radio Link Control}
\newacronym{rlf}{RLF}{Radio Link Failure}
\newacronym{rlnc}{RLNC}{Random Linear Network Coding}
\newacronym{rmr}{RMR}{RIC Message Router}
\newacronym{rmse}{RMSE}{Root Mean Squared Error}
\newacronym{rnis}{RNIS}{Radio Network Information Service}
\newacronym{rr}{RR}{Round Robin}
\newacronym{rrc}{RRC}{Radio Resource Control}
\newacronym{rrm}{RRM}{Radio Resource Management}
\newacronym{rru}{RRU}{Remote Radio Unit}
\newacronym{rs}{RS}{Remote Server}
\newacronym{rsrp}{RSRP}{Reference Signal Received Power}
\newacronym{rsrq}{RSRQ}{Reference Signal Received Quality}
\newacronym{rss}{RSS}{Received Signal Strength}
\newacronym{rssi}{RSSI}{Received Signal Strength Indicator}
\newacronym{rt}{RT}{Ray Tracer}
\newacronym{rtt}{RTT}{Round Trip Time}
\newacronym{ru}{RU}{Radio Unit}
\newacronym{rw}{RW}{Receive Window}
\newacronym{rx}{RX}{Receiver}
\newacronym{s1ap}{S1AP}{S1 Application Protocol}
\newacronym{sa}{SA}{standalone}
\newacronym{sack}{SACK}{Selective Acknowledgment}
\newacronym{sap}{SAP}{Service Access Point}
\newacronym{sc2}{SC2}{Spectrum Collaboration Challenge}
\newacronym{scef}{SCEF}{Service Capability Exposure Function}
\newacronym{sch}{SCH}{Secondary Cell Handover}
\newacronym{scoot}{SCOOT}{Split Cycle Offset Optimization Technique}
\newacronym{sctp}{SCTP}{Stream Control Transmission Protocol}
\newacronym{sdap}{SDAP}{Service Data Adaptation Protocol}
\newacronym{sdk}{SDK}{Software Development Kit}
\newacronym{sdm}{SDM}{Space Division Multiplexing}
\newacronym{sdma}{SDMA}{Spatial Division Multiple Access}
\newacronym{sdn}{SDN}{Software-defined Networking}
\newacronym{sdr}{SDR}{Software-defined Radio}
\newacronym{seba}{SEBA}{SDN-Enabled Broadband Access}
\newacronym{sgsn}{SGSN}{Serving GPRS Support Node}
\newacronym{sgw}{SGW}{Service Gateway}
\newacronym{si}{SI}{Study Item}
\newacronym{sib}{SIB}{Secondary Information Block}
\newacronym{sinr}{SINR}{Signal to Interference plus Noise Ratio}
\newacronym{sip}{SIP}{Session Initiation Protocol}
\newacronym{siso}{SISO}{Single Input, Single Output}
\newacronym{sla}{SLA}{Service Level Agreement}
\newacronym{sm}{SM}{Service Model}
\newacronym{smf}{SMF}{Session Management Function}
\newacronym{smo}{SMO}{Service Management and Orchestration}
\newacronym{sms}{SMS}{Short Message Service}
\newacronym{smsgmsc}{SMS-GMSC}{\gls{sms}-Gateway}
\newacronym{snr}{SNR}{Signal-to-Noise-Ratio}
\newacronym{son}{SON}{Self-Organizing Network}
\newacronym{sptcp}{SPTCP}{Single Path TCP}
\newacronym{srb}{SRB}{Service Radio Bearer}
\newacronym{srn}{SRN}{Standard Radio Node}
\newacronym{srs}{SRS}{Sounding Reference Signal}
\newacronym{ss}{SS}{Synchronization Signal}
\newacronym{sss}{SSS}{Secondary Synchronization Signal}
\newacronym{st}{ST}{Spanning Tree}
\newacronym{svc}{SVC}{Scalable Video Coding}
\newacronym{tb}{TB}{Transport Block}
\newacronym{tcp}{TCP}{Transmission Control Protocol}
\newacronym{tdd}{TDD}{Time Division Duplexing}
\newacronym{tdl}{TDL}{Tapped Delay Line}
\newacronym{tdm}{TDM}{Time Division Multiplexing}
\newacronym{tdma}{TDMA}{Time Division Multiple Access}
\newacronym{tfl}{TfL}{Transport for London}
\newacronym{tfrc}{TFRC}{TCP-Friendly Rate Control}
\newacronym{tft}{TFT}{Traffic Flow Template}
\newacronym{tgen}{TGEN}{Traffic Generator}
\newacronym{tip}{TIP}{Telecom Infra Project}
\newacronym{tm}{TM}{Transparent Mode}
\newacronym{to}{TO}{Telco Operator}
\newacronym{tr}{TR}{Technical Report}
\newacronym{trp}{TRP}{Transmitter Receiver Pair}
\newacronym{ts}{TS}{Technical Specification}
\newacronym{tti}{TTI}{Transmission Time Interval}
\newacronym{ttt}{TTT}{Time-to-Trigger}
\newacronym{tx}{TX}{Transmitter}
\newacronym{uas}{UAS}{Unmanned Aerial System}
\newacronym{uav}{UAV}{Unmanned Aerial Vehicle}
\newacronym{udm}{UDM}{Unified Data Management}
\newacronym{udp}{UDP}{User Datagram Protocol}
\newacronym{udr}{UDR}{Unified Data Repository}
\newacronym{ue}{UE}{User Equipment}
\newacronym{uhd}{UHD}{\gls{usrp} Hardware Driver}
\newacronym{ul}{UL}{Uplink}
\newacronym{um}{UM}{Unacknowledged Mode}
\newacronym{uma}{UMa}{Urban Macro}
\newacronym{umi}{UMi}{Urban Micro}
\newacronym{uml}{UML}{Unified Modeling Language}
\newacronym{upa}{UPA}{Uniform Planar Array}
\newacronym{upf}{UPF}{User Plane Function}
\newacronym{urllc}{URLLC}{Ultra Reliable and Low Latency Communications}
\newacronym{usa}{U.S.}{United States}
\newacronym{usim}{USIM}{Universal Subscriber Identity Module}
\newacronym{usrp}{USRP}{Universal Software Radio Peripheral}
\newacronym{utc}{UTC}{Urban Traffic Control}
\newacronym{vim}{VIM}{Virtualization Infrastructure Manager}
\newacronym{vm}{VM}{Virtual Machine}
\newacronym{vnf}{VNF}{Virtual Network Function}
\newacronym{volte}{VoLTE}{Voice over \gls{lte}}
\newacronym{voltha}{VOLTHA}{Virtual OLT HArdware Abstraction}
\newacronym{vr}{VR}{Virtual Reality}
\newacronym{vran}{vRAN}{Virtualized \gls{ran}}
\newacronym{vss}{VSS}{Video Streaming Server}
\newacronym{wbf}{WBF}{Wired Bias Function}
\newacronym{wf}{WF}{Waterfilling}
\newacronym{wg}{WG}{Working Group}
\newacronym{wi}{WI}{Wireless InSite}
\newacronym{wlan}{WLAN}{Wireless Local Area Network}
\newacronym{osm}{OSM}{Open Source \gls{nfv} Management and Orchestration}
\newacronym{pnf}{PNF}{Physical Network Function}
\newacronym{mtc}{MTC}{Machine-type Communications}
\newacronym{mns}{MnS}{Management Services}
\newacronym{ves}{VES}{\gls{vnf} Event Stream}
\newacronym{ei}{EI}{Enrichment Information}
\newacronym{fh}{FH}{Fronthaul}
\newacronym{fft}{FFT}{Fast Fourier Transform}
\newacronym{laa}{LAA}{Licensed-Assisted Access}
\newacronym{plfs}{PLFS}{Physical Layer Frequency Signals}
\newacronym{ptp}{PTP}{Precision Time Protocol}
\newacronym{cbrs}{CBRS}{Citizen Broadband Radio Service}
\newacronym{otic}{OTIC}{Open Testing and Integration Center}
\newacronym{sba}{SBA}{Service-Based Architecture}
\newacronym{cif}{CI}{cyberinfrastructure}
\newacronym{sonic}{SONiC}{Software for Open Networking in the Cloud}
\newacronym{ocp}{OCP}{Open Compute Project}
\newacronym{snmp}{SNMP}{Simple Network Management Protocol}
\newacronym{raid}{RAID}{redundant array of independent disks}
\newacronym{nfs}{NFS}{Network File Storage}
\newacronym{ci}{CI}{Continuous Integration}
\newacronym{cd}{CD}{Continuous Deployment}
\newacronym{dtn}{DTN}{Data Transfer Node}

\newacronym{dns}{DNS}{Domain Name Service}
\newacronym{nrpe}{NRPE}{Nagios Remote Plugin Executor}
\newacronym{ldap}{LDAP}{Lightweight Directory Access Protocol}
\newacronym{lan}{LAN}{Local Area Network}
\newacronym{vlan}{VLAN}{Virtual LAN}

\newacronym{ipmi}{IPMI}{Intelligent Platform Management Interface}
\newacronym{tor}{ToR}{Top-of-the-Rack}
\newacronym{lmn}{LMN}{Large Memory Node}
\newacronym{bgp}{BGP}{Border Gateway Protocol}
\newacronym{dhcp}{DHCP}{Dynamic Host Configuration Protocol}
\newacronym{vrf}{VRF}{Virtual Routing and Forwarding}
\newacronym{vpn}{VPN}{Virtual Private Network}
\newacronym{rma}{RMa}{Rural Macro}
\newacronym{hpc}{HPC}{High Performance Compute}

\newacronym{nu}{NU}{Northeastern University}
\newacronym{asic}{ASIC}{Application-specific Integrated Circuit}
\newacronym{rdma}{RDMA}{Remote Direct Memory Access}
\newacronym{roce}{RoCE}{RDMA over Converged Ethernet}
\newacronym{ovs}{OVS}{Open vSwitch}
\newacronym{frr}{FRR}{Free Range Routing}
\newacronym{ups}{UPS}{Uninterruptible Power Supply}

\newacronym{ntia}{NTIA}{National Telecommunications and Information Administration}
\newacronym{pii}{PII}{Personal and Identifiable Information}
\newacronym{irb}{IRB}{Institutional Review Board}
\newacronym{doi}{DOI}{Digital Object Identifier}

\newacronym{sdo}{SDO}{Standard-Development Organization}
\newacronym{ose}{OSE}{Open Source Ecosystem}
\newacronym{osc}{OSC}{O-RAN Software Community}
\newacronym{dop}{DOP}{Director of Operations}
\newacronym{pm}{PM}{Program Manager}
\newacronym{excom}{EXCOM}{Executive Committee}
\newacronym{iiot}{IIoT}{Industrial \gls{iot}}
\newacronym{lf}{LF}{Linux Foundation}

\newacronym{wiot}{WIoT}{Institute for the Wireless Internet of Things}
\newacronym{rl}{RL}{Reinforcement Learning}
\newacronym{drl}{DRL}{Deep Reinforcement Learning}

\newacronym{nofo}{NOFO}{Notice of Funding Opportunity}

\newacronym{onr}{ONR}{Office of Naval Research}
\newacronym{afosr}{AFOSR}{Air Force Office of Scientific Research}
\newacronym{afrl}{AFRL}{Air Force Research Laboratory}
\newacronym{arl}{ARL}{Army Research Laboratory}

\newacronym{arc}{ARC}{Aerial Research Cloud}
\newacronym{cast}{CaST}{Channel emulation scenario generator and Sounder Toolchain}

\newacronym{mno}{MNO}{Mobile Network Operator}
\newacronym{ct}{CT}{Continuous Testing}
\newacronym{oci}{OCI}{Open Container Initiative}

\newacronym{xai}{XAI}{Explainable AI}
\newacronym{sas}{SAS}{Spectrum Access System}

\newacronym{rem}{REM}{Random Ensemble Mixture}
\newacronym{ns3}{ns-3}{Network Simulator 3}

\newacronym{fcu}{FCU}{Flight Control Unit}
\newacronym{ros}{ROS}{Robot Operating System}
\newacronym{c2}{C2}{Command and Control}
\newacronym{esc}{ESC}{Electronic Speed Controller}
\newacronym{blos}{BLOS}{Beyond-Line-of-Sight}
\newacronym{bwp}{BWP}{Bandwidth Part}
\newacronym{paa}{PAA}{Phased-array Antenna}
\newacronym{ofdma}{OFDMA}{Orthogonal Frequency-Division Multiple Access}
\newacronym{mpc}{MPC}{Multipath Component}
\newacronym{qd}{Q-D}{Quasi-Deterministic}
\newacronym{sbr}{SBR}{Shooting-and-Bouncing Rays}
\newacronym{moi}{MOI}{Method of Images}
\newacronym{aoa}{AoA}{Angle of Arrival}
\newacronym{aod}{AoD}{Angle of Departure}
\newacronym{nyu}{NYU}{New York University}
\newacronym{csv}{CSV}{Comma-Separated Values}
\begin{document}
\title{Enabling Site-Specific Cellular Network Simulation \\ Through Ray-Tracing-Driven ns-3\vspace{-.3cm}

\thanks{This work was partially supported by the National Telecommunications and Information Administration (NTIA)'s Public Wireless Supply Chain Innovation Fund (PWSCIF) under Award No. 25-60-IF011, and by the National Science Foundation (NSF) under Award No. CNS 2112471.}
}


 \author{\IEEEauthorblockN{Tanguy Ropitault\IEEEauthorrefmark{1}\IEEEauthorrefmark{2}, Matteo Bordin\IEEEauthorrefmark{4}, Paolo Testolina\IEEEauthorrefmark{4}, \\ Michele Polese\IEEEauthorrefmark{4}, Pedram Johari\IEEEauthorrefmark{4}, Nada Golmie\IEEEauthorrefmark{3}, Tommaso Melodia\IEEEauthorrefmark{4}}
\IEEEauthorblockA{\IEEEauthorrefmark{1}\textit{Associate, CTL, National Institute of Standards and Technology, Gaithersburg, MD, USA}}
\IEEEauthorblockA{\IEEEauthorrefmark{2}\textit{Prometheus Computing LLC, Bethesda, MD, USA}}
\IEEEauthorblockA{\IEEEauthorrefmark{4}\textit{Institute for the Wireless Internet of Things, Northeastern University, Boston, MA, USA}}
\IEEEauthorblockA{\IEEEauthorrefmark{3}\textit{CTL, National Institute of Standards and Technology, Gaithersburg, MD, USA}}

 \{{tanguy.ropitault,nada.golmie\}@nist.gov}; \{{bordin.m, p.testolina, m.polese, p.johari, t.melodia\}@northeastern.edu}\vspace{-.4cm}
}

\maketitle


\begin{abstract}
Evaluating cellular systems, from \gls{5g} \gls{nr} and \gls{5g}-Advanced to \gls{6g}, is challenging because the performance emerges from the tight coupling of propagation, beam management, scheduling, and higher–layer interactions.  System-level simulation is therefore indispensable, yet the vast majority of studies rely on the
statistical \gls{3gpp} channel models.
These are well suited to capture average behavior across many statistical realizations, but cannot reproduce site-specific phenomena such as corner diffraction, street-canyon blockage, or deterministic line-of-sight conditions and angle-of-departure/arrival relationships that drive directional links.

This paper extends
5G\nobreakdash-LENA, an NR module for the system-level~\mbox{\gls{ns3}},
with a trace-based channel model that processes the \glspl{mpc} obtained from external ray-tracers (e.g., Sionna~\gls{rt}) or measurement campaigns.
Our module constructs frequency-domain channel matrices, and feeds them to the existing \gls{phy}/\gls{mac} stack without any further modifications. 
The result is a geometry-based channel model that remains fully compatible with the standard \gls{3gpp} implementation in \gls{5g}-LENA, while delivering site-specific geometric fidelity.
This new module provides a key building block toward \gls{dt} capabilities by offering realistic site-specific channel modeling, unlocking studies that require site awareness, including beam management, blockage mitigation, and environment-aware sensing.  
We demonstrate its capabilities for precise beam-steering validation and end-to-end metric analysis.
In both cases, the trace-driven engine exposes performance inflections that the statistical model does not exhibit, confirming its value
for high-fidelity system-level cellular networks research  and as a step toward  \gls{dt} applications.
\end{abstract}


\unnumberedfootnote{The mention of commercial products, their sources, or their use in connection with material reported herein is not to be construed as either an actual or implied endorsement of such products by the Department of Commerce.}
\unnumberedfootnote{U.S. Government work, not subject to U.S. Copyright}

\setlength{\belowcaptionskip}{-.3cm}

\glsresetall

\section{Introduction}%
System–level simulation is an indispensable tool for the design, evaluation, and optimization of \gls{5g} \gls{nr} networks.
Among the available simulators, the \gls{ns3} 5G-LENA module~\cite{nr-Lena} has established itself as a widely adopted open-source reference platform due to its compliance with \gls{3gpp} procedures and its support for configurable numerologies, bandwidth parts, and beamforming strategies.
In its default configuration, however, 5G-LENA relies on the statistical channel model of \gls{3gpp}~ \gls{tr} 38.901~\cite{3gpp.38.901,zugno2020implementation} (in the remainder of the paper, we refer to \gls{3gpp}~ \gls{tr} 38.901 as \gls{tr} 38.901 for brevity), in which large-scale parameters (e.g., path-loss, shadowing) and small-scale clusters (powers, delays, angles) are generated according to scenario-dependent probability distributions.  While this approach is well-suited for extensive Monte-Carlo studies and standardization activities, it cannot capture site-specific propagation features, e.g., deterministic blockage or dominant paths that drive the performance of highly directional links.
In particular, statistical clusters blur the geometric relationship between the transmitter and the receiver, masking the effects of \gls{los} occlusion and limiting the realism of beam-management or sensing algorithms that depend on accurate \gls{aod} and \gls{aoa} information.  

To close this gap, we introduce a 5G-LENA extension that processes \gls{mpc} traces, allowing ns-3 to compute a site-specific, physically realistic channel matrix through its native pipeline.
The proposed \emph{trace-based channel model} can operate with traces from ray-tracers (e.g., Sionna~\gls{rt}~\cite{hoydis2023sionna}, Remcom Wireless Insite, etc.) or measurement campaigns.
The new module replaces the stochastic cluster synthesis of \gls{tr} 38.901 with the delays, powers, phases, and angular information contained in the traces, while preserving 5G-LENA’s existing beamforming and scheduling block.
Consequently, users can perform site-specific evaluations of link adaptation, beam management and design procedures under the same \gls{mac} and \gls{phy} implementations already available in the simulator.  

The main contributions of this work are threefold.
First, we design and release\footnote{https://github.com/usnistgov/siolena} \project, an ns-3 module that builds frequency-domain channel matrices from ray traces, modeling location-specific multipath and Doppler effects.
Second, we maintain full compatibility with 5G-LENA release 4 without modifying its core scheduler, physical layer abstraction, or antenna models, thereby enabling a high-fidelity alternative to the native \gls{tr} 38.901 channel model. 
Third, we validate the module in two realistic \gls{3d} outdoor scenarios, 
showing that the proposed channel model reproduces the expected angular behavior of \gls{los} links and captures the pronounced end\nobreakdash-to\nobreakdash-end metrics degradation that occurs in \gls{nlos}, effects that the statistical model largely smooths out.  

Our \gls{ns3} extension, by integrating detailed propagation traces with a \gls{3gpp}-compliant network stack, enables reproducible research in advanced beamforming, positioning, and sensing in complex environments, and supports \gls{ai}/\gls{ml} data collection for wireless networks.
Because realistic propagation modeling is crucial for \gls{dt} use cases, integrating it with ns‑3’s system‑level simulator, which offers  5G core capabilities, marks a substantial stride toward practical \gls{dt} realization.

The remainder of the paper is organized as follows.
Section~\ref{SoA} reviews the state of the art and related works, highlighting our contributions.
In Section~\ref{sec:raytracing_channel}, we offer a brief overview of the 5G-LENA NR module and describe our trace‐based channel model extension.
Section~\ref{sec:eval} presents the beamforming validation results in the Place de l’Étoile scenario, analyzes end‐to‐end metrics in a Boston street canyon scenario,
and discusses the implications of trace‐based versus statistical modeling.
Finally, Section~\ref{sec:conclusion} concludes the paper and outlines directions for future work.



\section{Related Works} \label{SoA}
Stochastic channel models~\cite{3gpp.38.901,rappaport2017investigation,zugno2020implementation} are extremely effective in capturing general channel properties, 
but fail to reliably capture the propagation characteristics of specific environments.
In this context, \glspl{rt} enable physically consistent, site-specific full-stack simulations~\cite{lecci2021accuracy}, at the cost of a higher computational complexity. This shift toward realistic modeling aligns with the \gls{dt} paradigm, where virtual representations aim to accurately mirror the physical network's behavior and dynamics.
Among early efforts, \cite{assasa2019high} integrated a \gls{qd} channel model~\cite{gentile2018quasi} into the IEEE 802.11ad/ay \gls{ns3} implementation, providing for the first time trace-based channel generation capability to \gls{ns3}.
The \gls{qd} methodology, specifically designed to accurately model IEEE 802.11ad/ay channels, generates \glspl{mpc} based on the propagation conditions in a specific \gls{3d} environment, combining a deterministic ray-tracing component (large-scale fading) with a stochastic process (small-scale fading) whose distribution was extracted from an extensive measurement campaign~\cite{gentile2018quasi}.
The authors of~\cite{lecci2021accuracy} added similar capabilities to the \gls{5g} \gls{mmwave} module.

More recently, \glspl{rt} that exploit the parallelization capabilities of \glspl{gpu}, e.g., Sionna \gls{rt}~\cite{hoydis2023sionna}, dramatically reduced the channel computation time.
VaN3Twin framework \cite{milan-ns3Sionna} embeds Sionna \gls{rt} into the ms-van3t vehicular simulator via a modular UDP client-server architecture. Unlike prior \gls{ns3} integrations, VaN3Twin supports multi-RAT simulations across \gls{ieee} 802.11p, \gls{lte}-V2X, and NR-V2X, while enabling high-fidelity modeling of static and dynamic objects through \gls{3d} meshes. 
Ns3Sionna~\cite{berlin-ns3Sionna} introduces a client-server architecture based on ZeroMQ and protocol buffers to interface \gls{ns3} with Sionna \gls{rt}.

The integration with \gls{rt} allows reproducing a space and time-consistent, site-specific communication channel.
Namely,
    (i) \gls{los}/\gls{nlos} conditions are deterministically set by the node positions in the \gls{3d} environment;
    (ii) the overall channel gain is computed based on the interactions of the signal with the \gls{3d} environment model across its path(s) from the transmitter to the receiver; and
    (iii) correspondingly, the \gls{aoa} and \gls{aod} of the \glspl{mpc} are consistently determined by the aforementioned interactions.
We highlight that \emph{the angular information plays a key role particularly in directional communication scenarios}, i.e., with directional antennas or multi-element antenna arrays, as it is needed to determine the \gls{mpc} attenuation due to the antenna pattern and the beamforming vectors.
However, \cite{milan-ns3Sionna,berlin-ns3Sionna} reduce the channel response to an overall, pre-combined scalar gain, delay, and channel condition (\gls{los}/\gls{nlos}), discarding the \gls{aoa} and \gls{aod} information.
Customized antenna patterns and beamforming strategies can still be used, but they have to be implemented in Sionna, which disconnects their implementation from the \gls{ns3} antenna and beam management stack.
In practice, the standard \gls{5g} \gls{nr} beam management implementation, e.g., in the LENA module, is not compatible with these frameworks, and has to be re-implemented to interface with Sionna directly; or the simulation of directional antennas can not be controlled from \gls{ns3}, but has to be disabled in \gls{ns3} and externally managed in Sionna.
We believe that these elements limit the applicability of the aforementioned approaches, particularly for research in \gls{mimo} and \gls{fr2}, \gls{fr3}, and sub-THz frequency bands, where directionality is necessary to overcome the path loss.

In contrast, our architecture generates the \textbf{full spatial channel matrix} between each antenna element at runtime directly in \gls{ns3}, supporting configurable \gls{paa}, based on the complete \gls{mpc} information obtained from a \gls{rt} or a channel sounding measurement campaign, and feeding it into the \gls{ns3} pipeline.
This enables directional transmission to be modeled and beamforming vectors to be generated at runtime by \gls{ns3}, taking full advantage of its capabilities for antenna and beam management, unlocking its potential for beamforming and \gls{mimo} research.

\section{Raytracing-based Channel Model}
\label{sec:raytracing_channel}

\subsection{ns-3 LENA \gls{nr} Module and 3GPP Stochastic Model}
5G-LENA, the \gls{ns3} \gls{nr} module~\cite{nr-Lena}, supports 3GPP Release-15 end-to-end system-level simulations with support for flexible frame structures via multiple numerologies, \glspl{bwp}, and both time- and frequency-domain scheduling with dynamic \gls{tdd}, \gls{ofdma}, \gls{harq}, and \gls{mcs} adaptation. 
The \gls{phy} is abstracted through Exponential Effective \gls{sinr} Mapping (EESM) with lookup tables derived from a \gls{nr}-compliant link-level simulator. Additionally, the \gls{nr} module offers both ideal and realistic beamforming functionalities. Ideal methods rely on perfect channel or positional knowledge (e.g., cell-scan or direction-of-arrival), while realistic methods rely on \gls{srs} \gls{sinr}/\gls{snr} measurements to determine BF vectors.  
The 5G-LENA module
in \gls{ns3} provides a \gls{tr} 38.901 compliant channel modeling pipeline~\cite{zugno2020implementation} that captures large- and small-scale propagation effects for \gls{3gpp} \gls{umi}, \gls{uma}, and \gls{rma} scenarios. It is based on four classes as depicted in Fig.~\ref{fig:implementation}, top:

\begin{itemize}
  \item \texttt{ChannelConditionModel} determines the probabilistic channel condition, \gls{los}, \gls{nlos}, or \gls{o2i}, based on the 
  statistical models in \gls{tr} 38.901.
  The scenario parameters directly influence the \gls{los} probability, while the determined condition subsequently drives all path-loss and fading computations.
  \item \texttt{ThreeGppPropagationLossModel} implements large-scale fading mechanisms including path loss and shadow fading. Based on scenario configuration and \gls{los} condition, 
  it selects and applies the appropriate scenario-specific path-loss formula from Section 7 of \gls{tr} 38.901, incorporating frequency-dependent corrections, antenna height effects, and blockage losses to generate realistic link budgets.
  \item \texttt{ThreeGppChannelModel} generates the complex channel matrix using the cluster-based MPC modeling according to~\cite{3gpp.38.901}. Based on the scenario and channel condition, it determines cluster characteristics (number, powers, delays, angular spreads), generates MPC parameters, and constructs the spatial channel matrix accounting for antenna array geometry and spatial correlation. 
   \texttt{ThreeGppSpectrumPropagationLossModel} applies frequency-selective small-scale fading and beamforming processing to the transmitted signal spectrum. This class computes and applies the Doppler frequency shifts and the beamforming gain, and integrates the complete channel response with the signal spectrum to produce realistic instantaneous \gls{sinr} variations for link adaptation and scheduling algorithms.
\end{itemize}


\subsection{\project Trace-based Channel Model}

\begin{figure}[t]
  \centering
  \includegraphics[width=\linewidth]{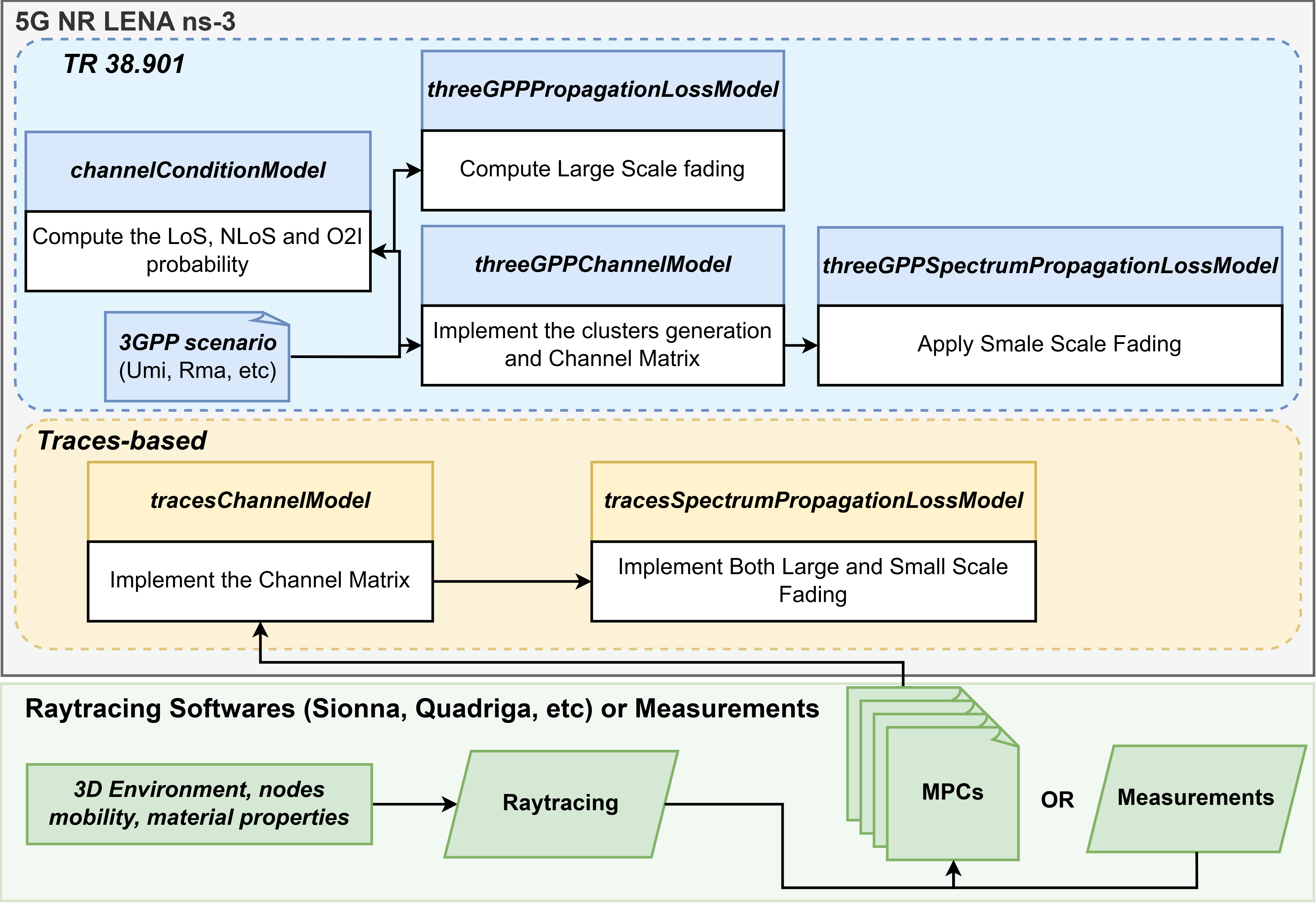}
  \caption{Overall channel model pipeline in 5G-LENA \gls{ns3}: (top) \gls{tr} 38.901–based classes, (bottom) newly added trace-based classes.}
  \label{fig:implementation}
      \vspace{-0.4cm}
\end{figure}

To complement the statistical \gls{3gpp} approach, we extended the LENA pipeline to utilize trace-based channel modeling. 
This approach enables to use \glspl{mpc} obtained from ray-tracers or measurement campaigns to enhance the channel model realism.
Two specialized classes (Fig.~\ref{fig:implementation}, bottom) operate in place of \texttt{ThreeGppChannelModel} and \texttt{ThreeGppSpectrumPropagationLossModel}:
\begin{itemize}
  \item \texttt{TracesChannelModel} processes trace-based precomputed \gls{mpc} from time-indexed \gls{csv} files generated by \gls{rt} software such as Sionna, or from measurement campaigns.
  Each trace entry specifies the propagation delay, the amplitude, phase, and both departure and arrival angles of an \gls{mpc}, for every transmitter–receiver link pair.
  For each \gls{mpc}, the antenna array response vectors are computed based on the provided angles, with appropriate phase shifts applied according to element positions and carrier wavelength.
  This process yields a frequency-domain channel matrix that faithfully represents the ray-traced or measured propagation environment.
  \item \texttt{TracesSpectrumPropagationLossModel} applies these deterministic channel matrices to the signal spectrum.
  Unlike the statistical approach, the Doppler shifts are computed directly from node velocities projected onto the \gls{mpc} propagation directions, providing physically accurate frequency shifts. Beamforming gains are calculated by multiplying the channel matrix by the transmit and receive steering vectors.
\end{itemize}

Since the \gls{rt}-derived or measurement campaigns \glspl{mpc} inherently encode the actual propagation conditions including \gls{los}/\gls{nlos} states, path delays, and received power levels, the \texttt{TracesChannelModel} eliminates the dependency on probabilistic modeling, capturing site-specific correlation and multipath structures.

\subsection{Sionna \gls{rt} Channel-Trace Generation for \gls{ns3}}
\label{subsection:SionnaRT}

While our trace-based channel model can use data from any \gls{rt} or measurement campaign that follows the required CSV format, in this article we showcase the workflow using \emph{Sionna RT}~\cite{hoydis2023sionna}. 
Sionna is an open-source Python library (TensorFlow-based) for link-level wireless simulation. Sionna RT extends it with GPU-accelerated 3D ray tracing that models reflection, diffraction and scattering under both LOS and NLOS conditions. 
To feed ns-3, we built a lightweight pipeline\footnote{\url{https://github.com/wineslab/sionna-channel-generator}} that  
(i) defines device trajectories (static, linear, circular, or user-provided),  
(ii) calls Sionna RT to compute the \glspl{mpc} (delays, complex gains, AoA, AoD, Doppler) for each Tx–Rx link, and  
(iii) exports these MPCs in the exact CSV schema expected by \texttt{TracesChannelModel} and \texttt{TracesSpectrumPropagationLossModel}.  
This exporter bridges Sionna RT’s detailed propagation output with our \project  ns-3 extension, enabling site-specific channel simulation without further manual conversion.


\section{Evaluation}
\label{sec:eval}
The goal of this section is twofold.  
First, we verify that the trace–based channel faithfully preserves the geometric information needed for beamforming, i.e., that the steering angles selected by 5G-LENA closely match the true \gls{los} directions embedded in the \gls{rt} trace.
Second, we show that this geometric fidelity propagates to end-to-end metrics: \gls{sinr}, \gls{mcs}, and user throughput.
In both case studies, we use the lightweight pipeline from Sec.~\ref{subsection:SionnaRT} to bridge Sionna RT and ns-3.  
Sionna RT generates the \glspl{mpc}, configured to include diffraction and up to four orders of specular reflection, from the 3D environment and node trajectories.  
Our pipeline then exports these MPCs into the  CSV schema expected by the SioLENA extension, enabling seamless import into ns-3.

The two scenarios intentionally involve a single \gls{gnb}–\gls{ue} pair and operate in the \gls{mmwave} band \gls{fr2} so that the results remain easily interpretable.
The proposed module, however, is agnostic to the number of nodes and to the carrier frequency: it can import traces for multiple \glspl{gnb} and \glspl{ue} simultaneously and works unchanged in FR1 or the emerging \gls{fr3} spectrum.

\subsection{Place de l'Étoile: Beamforming validation}
\begin{figure}[t!]
    \centering
    \begin{subfigure}[t]{0.45\linewidth}
    \includegraphics[width=\columnwidth]{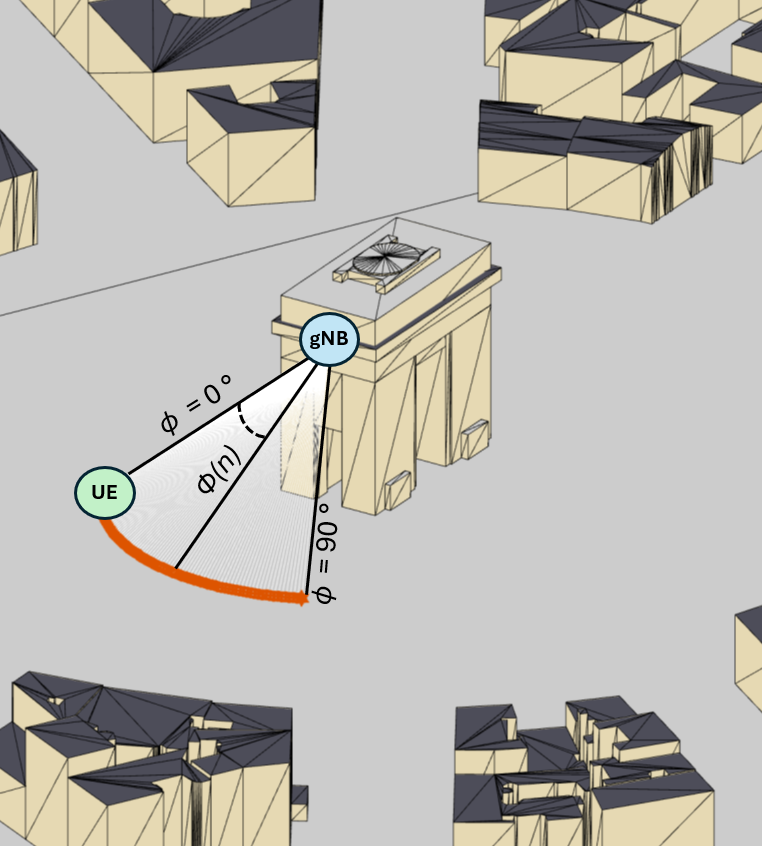}
    \setlength{\belowcaptionskip}{-.1cm}
        \vspace{-0.4cm}
  \caption{Visualization of the considered Étoile scenario. The angle $\phi$ denotes the azimuth AOD, with $\phi(n)$ representing the AOD at time step $n$. The UE trajectory is illustrated using orange markers.}
    \label{fig:etoile}
    \end{subfigure}
    \begin{subfigure}[t]{0.51\linewidth}
    \includegraphics[width=\columnwidth]{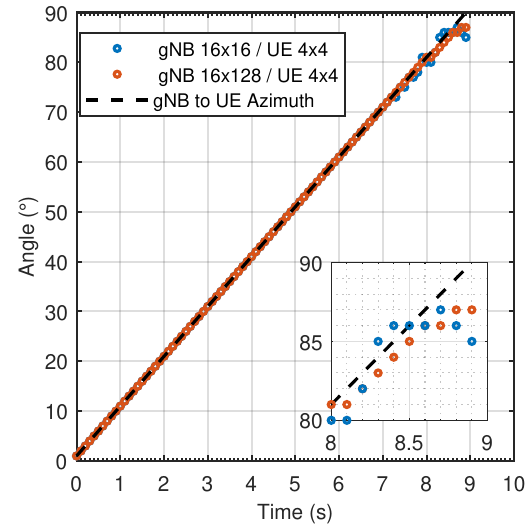}
    \setlength{\belowcaptionskip}{-.1cm}
    \vspace{-0.4cm}
    \caption{Time evolution of the steering azimuth selected for two \glspl{gnb} \glspl{paa}. An inset around 8-9~s highlights the increased deviation near the end-fire region.}
    \label{fig:az_time}
    \end{subfigure}
    \caption{Place de l’Étoile scenario.}
    \vspace{-0.4cm}
\end{figure}



We evaluate our trace-based channel model extension in a \gls{los} urban scenario at Place de l’Étoile, Paris. 
As depicted in Fig.~\ref{fig:etoile}, a single \gls{gnb}, operating at 28~GHz with 100~MHz of bandwidth, is mounted atop the Arc de Triomphe (10~m height), while a \gls{ue} at 1.5~m height moves along a circular trajectory, advancing by $1\,^\circ$ every 100~ms (i.e., the \gls{gnb}'s azimuth \gls{aod} sweeps from $0\,^\circ$ to $90\,^\circ$ over 9~s).

Beamforming training is performed every 100~ms using the \texttt{idealBeamforming} procedure from LENA 5G‐NR, configured to scan the azimuth plane in $1\,^\circ$ steps and the elevation plane in $10\,^\circ$ steps.
At each beam‐training interval, the \gls{gnb} selects the steering vector that yields the highest received power at the \gls{ue}.
We emphasize that \texttt{idealBeamforming} is idealized: beam‐training frames are not transmitted over the medium, so no interference arises during beam training.   Two \gls{gnb} array configurations are studied: $16\times16$ \gls{paa} and $16\times128$ \gls{paa} while the \gls{ue} uses a $4\times4$ \gls{paa}.
The \gls{paa} elements are positioned in the $y$–$z$ plane and the antenna elements are isotropic.



Fig.~\ref{fig:az_time} compares the azimuth
of the beamforming vector selected in \gls{ns3} (markers) with the true \gls{ue} azimuth (dashed line).
The two patterns are indistinguishable for most of the 9~s trajectory, confirming the correctness of the model.
The matching persists until the \gls{los} angle approaches about $70\,^\circ$, beyond which the steering angle approaches the end-fire region, as the \gls{paa} elements lie in the $y$–$z$ plane, and thus exhibit limited azimuthal resolution.
The inset highlights the increased beam spread between 8~s and 9~s, with the three largest errors occurring within this period: $5\,^\circ$ for the $16\times16/4\times4$ array and $3\,^\circ$ for the $16\times128/4\times4$ array.

The beam-steering for the $16\times16/4\times4$ configuration has a mean error of $-0.17\,^\circ$ and a \gls{rmse} of $0.74\,^\circ$, while the $16\times128/4\times4$ configuration achieves a mean error of $-0.14\,^\circ$ and an \gls{rmse} of $0.51\,^\circ$.
The lower \gls{rmse} of the latter array reflects its larger element count: more elements yield a narrower beamwidth and finer angular discrimination in azimuth, reducing quantization error and the magnitude of end-fire outliers.

These results confirm that (i) the trace-based channel model preserves directional information from the imported \glspl{mpc}, and (ii) \texttt{idealBeamforming} coupled with the trace-based channel model recovers the true \gls{los} direction in most cases.
The expected degradation near end-fire, and the reduced error when moving from $16\times16$ to $16\times128$, demonstrate that our framework can be used not only to study \gls{paa} design trade-offs, to evaluate beamforming and beam-training algorithms in realistic scenarios, but also to generate realistic, labeled datasets for \gls{ml}-based beam-steering or channel estimation algorithms.
Such \gls{ml}-driven methods require training data with geometrically consistent steering directions, which cannot be obtained using the \gls{3gpp} statistical channel model.
\subsection{Boston Street Canyon: End-to-End Analysis}

\begin{figure}[t!]
    \centering
    \includegraphics[width=\linewidth]{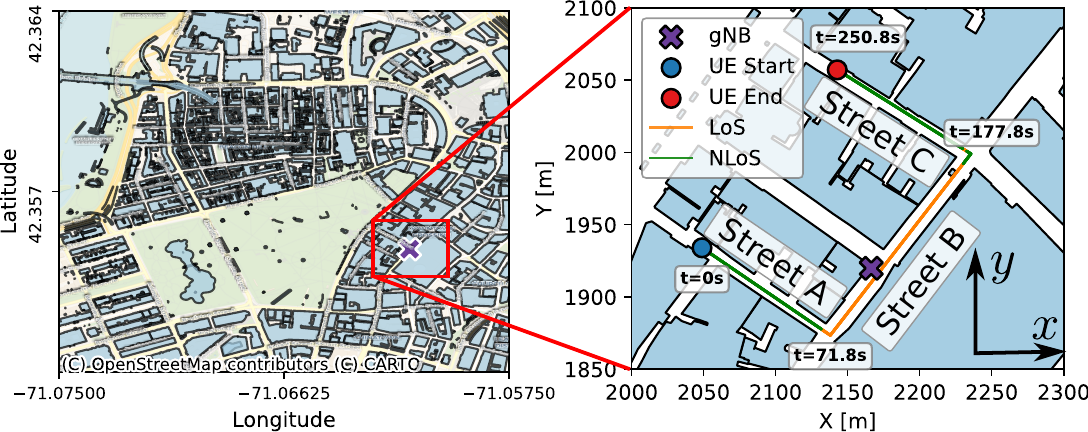}
        \setlength{\abovecaptionskip}{-.1cm}
    \setlength{\belowcaptionskip}{-.6cm}
    \vspace{-0.2cm}
    \caption{Boston simulation scenario.}
    \label{fig:SimulationScenario}
\end{figure}

This validation quantifies the end-to-end impact of our trace-based channel model. 
Leveraging the Boston digital-twin street-canyon model of~\cite{Boston_twin}, we deploy a single \gls{gnb} at 10~m height and let a pedestrian \gls{ue} at 1.5~m height walk 375~m in 250~s at 1.5~m/s.  
The route crosses three consecutive streets, \emph{A}, \emph{B}, and \emph{C}, and  alternates between \gls{los} and \gls{nlos} visibility: Streets~A and~C are \gls{nlos}, whereas Street~B is predominantly \gls{los} with a brief blockage near its end (see Fig.~\ref{fig:SimulationScenario}).

We compare the results obtained using our trace-based channel model, based on the \glspl{mpc} generated in Sionna, against those obtained with the \gls{umi} street-canyon \gls{3gpp} channel model.
The \gls{gnb} employs a $16\times16$ \gls{paa}, whereas the \gls{ue} is equipped with a $4\times4$ \gls{paa}.  
Downlink traffic is generated at a constant \gls{udp} rate of \SI{10000}{packets\per\second} with \SI{1500}{byte} packets, corresponding to 122~Mb/s; the \gls{gnb} buffer stores up to \SI{10000}{packets}.  
Beam training is performed every \SI{100}{\milli\second} using a $10\,^\circ$ scan grid, and the 3GPP \gls{los}/\gls{nlos} condition is re-evaluated at the same \SI{100}{\milli\second} interval.

\begin{figure}[t]
    \centering
    \vspace{-.3cm}
    \begin{subfigure}[t]{0.87\linewidth}
        \centering
        \includegraphics[width=\linewidth]{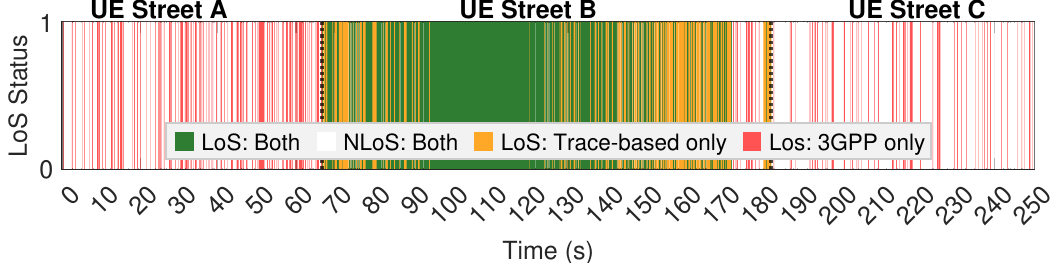}
        \setlength{\abovecaptionskip}{-.3cm}

        \setlength{\belowcaptionskip}{-.1cm}
        \caption{Comparison of \gls{los} status between the trace‐based channel model and \gls{tr}~38.901 (\gls{3gpp}). 
}
        \label{fig:boston_los}
    \end{subfigure}\par\medskip
    \begin{subfigure}[t]{\linewidth}
        \setlength{\fwidth}{\columnwidth}
        \setlength{\fheight}{.7\columnwidth}
        \centering
        \includegraphics[width=0.88\linewidth]{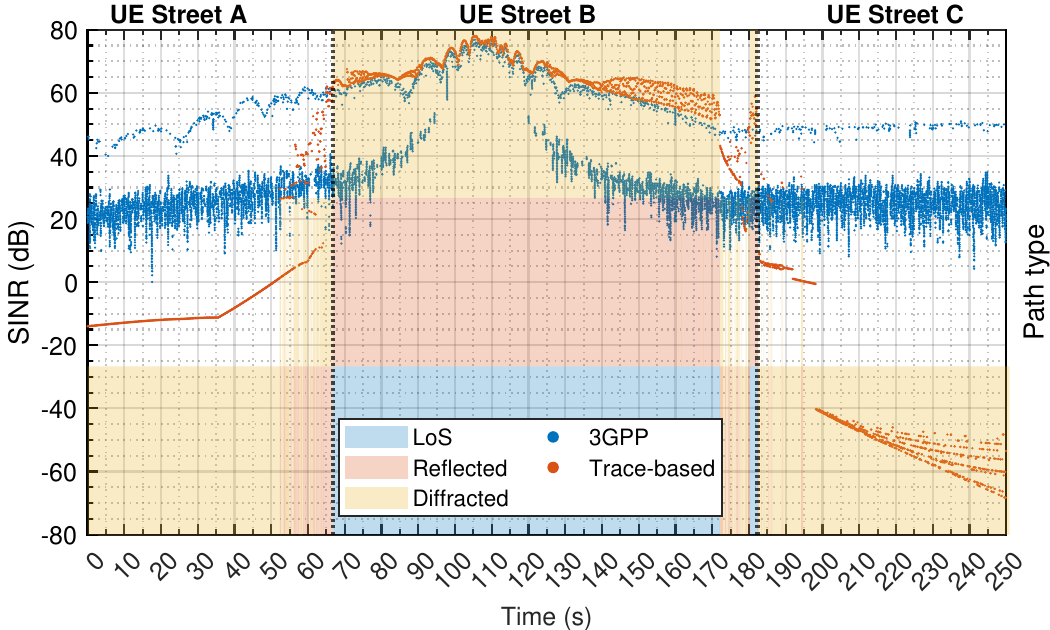}
        \setlength{\belowcaptionskip}{-.1cm}
        \vspace{-0.2cm}
        \caption{Instantaneous downlink \gls{sinr} for the two channel models (red markers trace-based and blue markers \gls{3gpp}). Coloured backgrounds mark the type of \glspl{mpc} generated by Sionna (diffracted, reflected, or  \gls{los}).}
        \label{fig:boston_sinr}
    \end{subfigure}
    \caption{\gls{los} agreement and \gls{sinr} evolution versus time.}
    \label{fig:boston_los_sinr}
\end{figure}

\textbf{Channel Characterization.}
Fig.~\ref{fig:boston_los} shows the \gls{los} agreement between \gls{3gpp} and trace-based channel model, and Fig.~\ref{fig:boston_sinr} shows both the \glspl{mpc} (\gls{los}, reflected, diffracted) in the Sionna trace and the \gls{sinr} evolution for \gls{3gpp} and trace-based channel model.
We can identify three distinct regions: (1) \textit{Early \gls{nlos}} (0~s to 67~s): The UE is in Street~A (physically NLoS).  
    The trace-based model correctly remains NLoS, whereas \gls{tr}~38.901 erroneously reports LoS (red bars).  
    For the trace-based channel, we can observe that diffraction dominates, with specular reflections appearing only as the UE rounds the corner into Street~B (around 53~s).
(2) \textit{Predominantly \gls{los}} (67~s to 182~s): In Street~B the link is LoS except during a blockage from 172~s to 180~s (see Fig.~\ref{fig:SimulationScenario}).  
    Both models agree most of the time (green bars), though \gls{tr}~38.901 misclassifies $\sim32\,\%$ of samples as NLoS (orange bars).  
    The trace-based channel is driven by a strong LoS MPC (except during the blockage), with reflected and diffracted components.  
(3) \textit{Late \gls{nlos}} (182~s to 250~s): Upon entering Street~C, the physical link returns to NLoS.  
    The trace-based model remains strictly NLoS, while \gls{tr}~38.901 occasionally indicates LoS.  
    Again, for trace-based channel model, brief specular reflections punctuate an otherwise diffraction-dominated channel.

\begin{figure}[htbp]
  \centering

  \begin{subfigure}[t]{\columnwidth}
    \centering
    \includegraphics[width=0.90\linewidth]{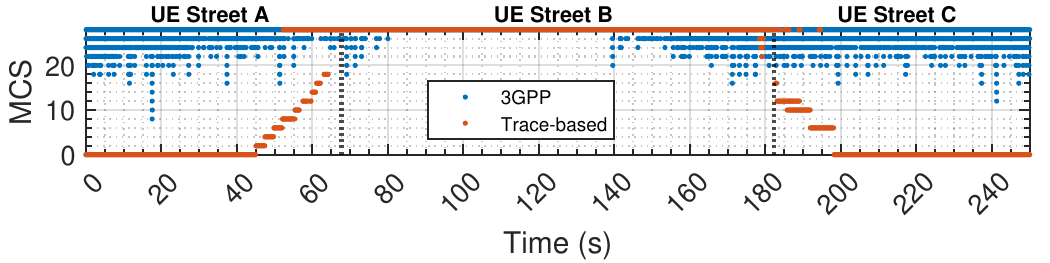}
    \setlength{\belowcaptionskip}{0cm}
    \caption{\gls{mcs} at the \gls{gnb}.}
    \label{fig:boston_mcs}
  \end{subfigure} 

  \begin{subfigure}[t]{\columnwidth}
    \centering
    \includegraphics[width=0.90\linewidth]{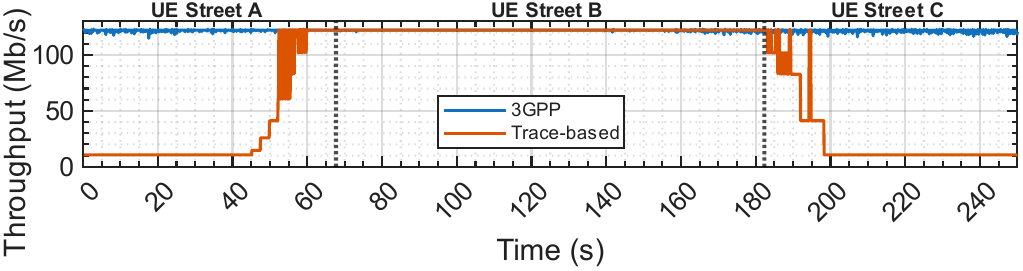}
    \setlength{\belowcaptionskip}{0cm}
    \caption{\gls{udp} throughput measured at the \gls{ue}.}
    \label{fig:boston_tp}
  \end{subfigure}

  \begin{subfigure}[t]{\columnwidth}
    \centering
    \includegraphics[width=0.90\linewidth]{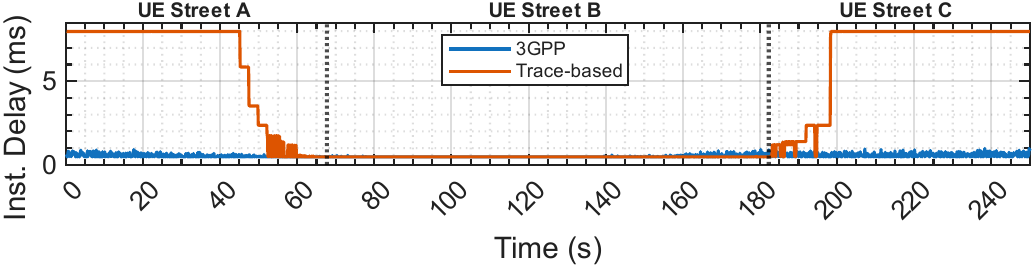}
    \setlength{\belowcaptionskip}{0cm}
    \vspace{-0.2cm}
    \caption{Packet delay experienced at the \gls{ue}.}
    \label{fig:boston_delay}
  \end{subfigure}

  \caption{Link adaptation, throughput, and delay under the two channel models.}
  \label{fig:boston_mcs_tp_delay}
  \vspace{-.2cm}
\end{figure}




\textbf{SINR.}
Fig.~\ref{fig:boston_sinr} shows the instantaneous downlink \gls{sinr} at the \gls{ue} for both channel models.  
This \gls{sinr} is governed by a tight interplay between the time-varying channel and the beamforming.  
As the \gls{ue} moves, some \glspl{mpc} appear while others fade; their relative phases determine whether they add constructively or destructively in the \gls{paa} plane.  
Periodic beam training then computes the weights that maximize the combined power at the array ports.  
Thus, the received power, and therefore the \gls{sinr}, is the joint outcome of the instantaneous \gls{mpc} set and the current beam-steering weights.
Two global trends emerge.  
(i) A switch from \gls{nlos} to \gls{los} visibility yields an immediate \(\sim\!20\) dB jump in \gls{sinr}.  
(ii) For both models the \gls{sinr} rises from 0 s to 106 s, while the \gls{ue} and \gls{gnb} approach their closest separation, and then falls  as they move apart.

A closer inspection of the trace-based \glspl{mpc} in Fig.~\ref{fig:boston_sinr} shows that the channel is dominated by diffracted paths in the intervals 0~s to 67~s and again after 182~s. In the former interval, a single diffracted ray keeps the \gls{sinr} stable around \(-15\)~dB, increasing slowly.
After 35~s, as the \gls{gnb} nears the corner of Street~A, the diffraction angle becomes more favorable: the \gls{aoa} of the diffraction path aligns more closely with the boresight of the \gls{ue}’s \gls{paa}, resulting in a sharper \gls{sinr} increase.
On the contrary, after 197~s, the \gls{sinr} sharply declines as the diffraction path’s \gls{aoa} steers the beam into the array's end-fire region, reducing significantly the azimuthal gain.
Between 53~s and 67~s, and 182~s and 195~s, intermittent specular reflections raise the \gls{sinr} by \(\sim\!20\)~dB.  
Once the \gls{ue} enters Street~B at 67~s and the \gls{los} is established, the \gls{sinr} climbs above 60~dB, peaks near 80~dB at minimum \gls{gnb}–\gls{ue} separation, then declines as the nodes move apart.  
A brief blockage from 172~s to 180~s removes \(\sim\!20\)~dB, with a short rebound from 180~s to 182~s before the \gls{ue} turns into Street~C.

The \gls{tr}~38.901 statistical model smooths these fluctuations: the \gls{sinr} stays mostly in the 10~dB to 45~dB range even during the initial \gls{nlos} interval, with frequent \gls{los} appearances, and rarely falls below 10~dB, 
due to distance-based cluster power distributions that do not reflect the site geometry.


%

\textbf{End-to-end Metrics.}
Fig.~\ref{fig:boston_mcs} shows how \gls{amc} adapts to the time‐varying \gls{sinr} of Fig.~\ref{fig:boston_sinr}. In 5G‐LENA, \gls{amc} selects each transport‐block’s \gls{mcs} to maximize spectral efficiency under a 10\,\% BLER constraint. During early NLoS when the \gls{ue} is in Street~A (0~s to 45~s), \gls{amc} picks \gls{mcs}~0 ($\text{\gls{sinr}}\sim-15$ dB), whereas \gls{tr}~38.901 reports ($\text{\gls{sinr}}\sim20$ dB) and thus \gls{mcs} 8 to 28 but mostly larger than 24. As the UE nears the Street~A–B corner (45~s to 50~s), the trace‐based \gls{sinr} rises and \gls{mcs} climbs to 8 to 28 (depending on reflections), while the 3GPP model remains at high indices. When the \gls{ue} enters Street~B (67~s to 182~s), LoS emerges and both models reach \gls{mcs} 28, though 3GPP intermittently dips to 16 due to the  NLoS condition; trace‐based stays at 28 except during blockage (down to 22). After the \gls{ue} entering Street~C (182~s to 250~s), trace‐based \gls{mcs} gradually falls to 0 as reflections vanish, whereas the statistical model selects \gls{mcs} 12 to 28.

The \gls{ue}’s throughput (Fig.~\ref{fig:boston_tp}) and delay (Fig.~\ref{fig:boston_delay}) mirror the \gls{mcs}: lower indices shrink transport‐block size and spectral efficiency, requiring more radio link control segments, extra HARQ rounds, and fewer bits per transmission time interval, thereby reducing throughput and raising delay.  
In the trace-based model, while the UE traverses Street~A under NLoS (0~s to 45~s), \gls{mcs} remains at 0, capping throughput at \(\sim\!10\)~Mb/s and inflating delay to \(\sim\!8\)~ms. Between 45~s and 67~s, as the UE rounds the Street~A–B corner, emerging specular reflections and a diffraction angle that aligns more closely with the \gls{ue}'s \gls{paa} boresight, drive \gls{mcs} up to 28, elevating throughput to 122~Mb/s and cutting delay below 0.5~ms.
 While traversing Street~B under sustained LoS (67~s to 172~s), throughput remains at 122~Mb/s with minimal delay; when the obstacle blocks LoS (172~s to 180~s), \gls{mcs} dips to 22 yet maintains \(\sim\!100\)~Mb/s and low delay. In Street~C, from 182~s to 197~s, throughput drops to \(\sim\!50\)~Mb/s and delay peaks at \(\sim\!2\)~ms, except when specular reflection punctually appears (186~s to 195~s), restoring full rate and low delay. Finally, at 197~s, the \gls{ue} is steering into its \gls{paa} end-fire region, forcing \gls{mcs} back to 0, reducing throughput to \(\sim\!10\)~Mb/s and returning delay to \(\sim\!8\)~ms. These results clearly capture the site-specific performance variations induced by channel evolution.

By contrast, the \gls{tr}~38.901 statistical street‐canyon model smooths localized SINR fluctuations, sustaining throughput near 122~Mb/s (mean 121.92~Mb/s) and limiting delay to 0.5–1~ms (mean 0.54~ms), and thus does not reflect these site‐specific effects.

\noindent\textbf{Discussion.} 
Deterministic \gls{mpc} traces are indispensable whenever the \emph{geometric} components of the channel, and not merely its average gain, drive system behavior (e.g. joint communication-sensing, and blockage prediction).
Moreover, statistical and trace-based models are complementary: the \gls{tr}~38.901 offers rapid Monte-Carlo sweeps for large network layouts, whereas the trace pinpoints worst-case corners, indoor–outdoor transitions, or moving blockers. 
Additionally, the runtime exposure of full channel matrices supports hybrid or learning-based beam selection \emph{without} re-running the ray-tracer.
Finally, as the trace parser is agnostic to the ray-tracing tool and to the carrier frequency, the approach extends to sub-THz industrial sensing and satellite backhaul where \gls{los} and specular reflections dominate.

\section{Conclusions}
\label{sec:conclusion}
This work integrates site-specific ray-tracing into the open-source \gls{5g}-LENA simulator while preserving its \gls{3gpp}-compliant beamforming, scheduling, and link-adaptation stack.
The trace-based channel model reproduces deterministic angular dynamics and, when compared with the standard \gls{tr}~38.901 implementation, reveals significant differences in beam-steering accuracy, \gls{sinr}, \gls{mcs} selection, and user throughput under blockage and corner diffraction.


Ongoing work targets more challenging scenarios: \emph{(i)} multi-site deployments with inter-cell interference, \emph{(ii)} spectrum sharing and coexistence, \emph{(iii)} mobility with multiple \glspl{gnb} and \glspl{ue} executing concurrent beam training and handover, and \emph{(iv)} dynamic blocker models and traffic heterogeneity. Future directions also include exploring how this realistic propagation modeling could contribute to \gls{dt} applications for network optimization and AI-driven network management.

\bibliographystyle{IEEEtran}
\bibliography{biblio}

\end{document}